# Innovation adoption: Broadcasting vs. Virality

**Yujia Zhai** *(Corresponding Author)*

*Department of Information Resource Management, School of Management, Tianjin Normal University, Tianjin, China;*

*School of Information Management, Wuhan University, Wuhan, Hubei, China.*

*Telephone:(86-22) 2376-6058*

*E-mail:* [zhaiyujiachn@gmail.com](mailto:zhaiyujiachn@gmail.com)

**Ying Ding**

*School of Information, University of Texas at Austin, TX, U.S.A.*

*E-mail:* [ying.ding@ischool.utexas.edu](mailto:ying.ding@ischool.utexas.edu)

**Hezhao Zhang**

*Department of Information Resource Management, School of Management, Tianjin Normal University, Tianjin, China.*

*E-mail:* [zhanghezhao794@gmail.com](mailto:zhanghezhao794@gmail.com)

# Abstract

Diffusion channels are critical to determining the adoption scale which leads to the ultimate impact of an innovation. The aim of this study is to develop an integrative understanding of the impact of two diffusion channels (i.e., broadcasting vs virality) on innovation adoption. Using citations of a series of classic algorithms and the time series of co-authorship as the footprints of their diffusion trajectories, we propose a novel method to analyze the intertwining relationships between broadcasting and virality in the innovation diffusion process. Our findings show that broadcasting and virality have similar diffusion power, but play different roles across diffusion stages. Broadcasting is more powerful in the early stages but may be gradually caught up or even surpassed by virality in the later period. Meanwhile, diffusion speed in virality is significantly faster than broadcasting and members from virality channels tend to adopt the same innovation repetitively.

# Innovation adoption: Broadcasting vs. Virality

## 1. Introduction

Innovation is what drives human civilization forward (Rogers, 2010; Perry-Smith & Mannucci, 2017). In the late 1920s, Alexander Fleming accidentally discovered a naturally ring-shaped mold free of staphylococcus bacteria in one of his samples. Classifying the mold as being from the genus Penicillium, he dubbed the substance "penicillin." However, the journey towards Penicillin's mainstream adoption did not happen overnight. Fleming's findings did not receive much attention at first. This is often attributed to his incompetence as a communicator. Although he continued to publish papers to further describe the characteristics of penicillin, his peers were not interested (Martin, 2007). It was not until 1940, more than a decade after the discovery of this miracle substance, that mass production of penicillin was proposed and promoted to American pharmaceutical companies. Five years later, in 1945, Fleming shared the Nobel Prize in Medicine with Howard Florey and Ernst Chain for their contribution to the research and promotion of penicillin in the years following its initial discovery (Sykes, 2001). Obviously, the journey of innovation adoption is crucial to its ultimate success. Most innovations, oftentimes serendipitous accidents, experienced extremely hardness to reach massive adoptions. From bank cards to nuclear power, the average time for their widespread takes 39 years (Hanna et al., 2015). The mystery of the tension that delays the adoption of innovative discoveries has spurred researchers to investigate adoption patterns of successful innovations to try and understand the key steps in spreading scientific discoveries at an appropriate rate.

Three key factors in understanding what hinders the widespread adoption of some innovations are access, evaluation, and promotion. When an innovation is first proposed, the number of people who can access related information, get interested, and gain trust in the innovation is limited. Also, every innovation is imperfect at the outset – it may be limited in scope, focusing only on solving a single problem, or its results might be unsatisfactory or inconvenient to use (Rogers, 2010). To overcome these obstacles, innovators must first explain their findings well. Second, early adopters must constantly evaluate the performance and quality of the innovation, expanding its influence and reputation while building trust (Larsen, 2011). Third, successful innovations rely on the efforts of innovators, supporters, and even opponents to constantly promote, modify, improve, and criticize their findings (Green et al., 2009). Therefore, in general, effective communication channels and follow-up behavior within the network of influencers is crucial to the diffusion of an innovation.

In this paper, we use broadcasting and virality to define diffusion channels from a structural perspective. Broadcasting is dominated by a large burst of adoptions from the source of innovation. An example of this is the practice of publishing articles in open-access journals, enabling anyone to read, download and cite the research. In contrast, virality can be seen as a time-varying adoption cascade that can be propagated from individual to individual through potentially multiple generations of adopters. Virality is driven by social contagion mechanisms and dominated by interpersonal communication, in which one's adoption of an innovation is a function of their exposure to others' knowledge, attitude, or behavior (Yue et al., 2019). Specifically, we use scientific collaboration to represent virality in the diffusion of academic innovation. We are curious as to how we can restore the diffusion channels of innovation, how to quantify their diffusion powers, and how different channels promote the diffusion of innovation in different stages. These questions can provoke insights into the social and psychological dimensions of this dynamic process.

This article is outlined as follows. First, we discussed the related works from the perspective of tracing innovation diffusion and studies on diffusion channels. We then describe the dataset and the methodology that we use in this article. Results are compared with existing related studies. Finally, conclusions and suggestions for future work are offered. Our results can therefore play a fundamental role in guiding and assisting policy-makers, funding bodies, and researchers.

## 2. Related work

### 2.1 Tracing innovation diffusion

Tracking the diffusion path of innovation needs to record (1) the adopters; (2) the adoption time; and (3) all information sources from which the adopters could conceivably have learned about the innovation (Goel et al., 2012). It is difficult to gather all of this information in reality since diffusion can last for years or even decades. Scientists often receive only a small portion of sample data through questionnaires and interviews.

Citations have long been considered the footprint of innovation diffusion.(Jaffe & Trajtenberg, 1996; Martens & Goodrum, 2006; Rong & Mei, 2013; Min et al., 2018; Zhai et al., 2018). Scientists have identified the knowledge contained in literature as an innovation entity, and the citation history of the literature is the path of diffusion. Regarding citations as the path of diffusion, Rong and Mei (2013) studied the competition and collaboration relationships among algorithms(innovations) in computer science. They found that the adoption rate of one innovation increases with the proportion of its competitors or collaborators adopted by the user. Zhai, Ding, and Wang (2018) built a paper-subject network from the citations of LDA. Using topic modeling to identify the role of LDA in its citations, they found that a scientific innovation can first diffuse to adjacent disciplines.

We consider scientists as innovation adopters and explore how innovation spreads among them. Specifically, we use academic collaboration as a diffusion path for scientific innovation, because the process of scientific collaboration is accompanied by the dissemination, sharing and exchanging of explicit and tacit knowledge among scientists (Eslami, Ebadi, & Schiffauerova, 2013). In addition to being documented as co-authorship in literature, the collaboration between scientists also exists as a strong and intimate social relationship. These kinds of relationships represent important channels for social contagion. Within the context of the growing complexity of research, collaboration has been considered one of the most crucial and common phenomena in the scientific community (Wuchty et al., 2007; Mukherjee et al., 2017). Previous research also indicated, in the process of knowledge dissemination and creation, that self-organization and autonomy of scientific research give the scientific collaboration network the ability to fit the dynamic model of knowledge diffusion most appropriately (Yang et al., 2015).

### 2.2 Broadcasting and virality

According to Rogers (2010), communication channels can be divided into two categories: broadcasting via mass media and virality via interpersonal connections. Mass communication mainly relies on mass media, such as television, radio, newspapers, magazines or books, to deliver information to a large number of anonymous and heterogeneous users. Interpersonal connections include all types of face-to-face communication between two or more individuals. From the view of tree traversal, broadcasting refers to a large-scale transmission event, in which a single source spreads content to a large number of people, and virality refers to a cascade of shared events each between a sender and their associates(Zhang et al., 2020).

Both of these diffusion channels have a direct and powerful influence on innovators who are actively seeking information on new technologies. As early as 1990, Brancheau and Wetherbe studied the dissemination of spreadsheet software in organizations and more than 500 employees from 18 large companies participated. Results showed that broadcasting is crucial to respondents first hearing about spreadsheet software and forming impressions. By studying the cascade of word-of-mouth communication interactions, Susarla et al. (2016) found that whether a new product can be designed to be viral depends on whether it can attract a large number of early adopters. With respect to broadcasting, researchers are more inclined to pay attention to virality effects. Zhang et al. (2016) reviewed the important studies in information diffusion, especially the results of information cascading effects. They pointed out that most of the research on information diffusion focused on viral spreading, while ignoring the influence of the broadcast mechanism. However, it was found that broadcasting was the dominant mechanism of information diffusion of a major health event on Twitter (Liang et al., 2019).

In the adoption decision process, virality is more important during the persuasion phase than during the knowledge stage (Rogers, 2010). At the same time, individuals who are more interpersonally connected within a social system are more likely to adopt an innovation than individuals who are less interconnected within the system. Therefore, virality can effectively eliminate "social-psychological barriers" and the reluctance or indifference of users when adopting new technologies (Nejad et al., 2014). As early as 2002, Lee and his collaborators found that highly intimate interpersonal interactions are critical to fostering initial trust in new innovations. This, in turn, leads to a high probability of adopting technological innovation during the later stages of the diffusion process, especially for imitators or later adopters. In addition to enhancing the willingness to adopt innovations, compared with the broadcast model, viral diffusion also increases the possibility of cross-ideological sharing and thus increases political diversity on social media(Liang, 2018).

Scientists have made great attempts to quantify virality to study the relationship between cascade and diffusion size. Most studies have relatively simple definitions of virality. When analyzing the dissemination of scientific literature, Guerini et al. (2012) defined virality as the volume of downloads, bookmarks, and citations an article receives. When studying the diffusion of meme in social media, Weng et al. (2013) used the popularity of the meme as an indicator of its virality, and predicted the meme virality based on the community concentration in the early stage of the diffusion. Similarly, Vougiouklis et al. (2020) measured the virality potential of tweets based on the count of retweets and favorites.

Further research on virality begun to consider not only the size of a given cascade tree, but also its depth and branches. Goel et al. (2015) proposed an indicator termed structural virality to quantify the intuitive difference between broadcasting and viral diffusion of news, videos, pictures and petitions on Twitter. The structural virality of the diffusion tree is calculated by the sum of the average path length between all nodes, where the more viral of a cascade tree, the larger the structural virality would be. However, the diffusion cascades are actually non-isomorphic, but Goel's indicator treats them as undirected graphs and fails to capture the root nodes, which hinders its ability to differentiate cascades. To eliminate this trap and more accurately quantify the virality of cascades, Zhang et al. (2020) proposed a root-based method called cascade virality, which calculates the average distance between a node and its descendants and sum the distances on all nodes.

The most relevant study of our research is Garas et al.'s work (2017) on the innovation diffusion in robotic surgery. They used citation networks to study the different diffusion stages through which innovation in surgery typically progresses and proposed several indicators to measure broadcasting and virality, including paper citations, cascade size, structural depth and width. Following this trajectory, we transform the citation network into a cascade of authors and try to study the intermingling of broadcasting and virality during the innovation adoption process. In this paper, we define the diffusion channel as where the researcher obtains an academic work and decides to cite. For scientific research, an innovation can be expressed as an academic article. Adoption can be seen as a citation to this article, and each author of the citation is an adopter.

## 3. Method

### 3.1 Dataset

In order to select some representative cases of innovation, we need to follow some criteria. First of all, an innovation should have been proposed for a period time and received enough citations to restore a typical diffusion trajectory. Second, these innovations should be adopted by virous disciplines and applied to answer different research questions. Finally, there should be both similarities and diversity between these cases, so we can obtain more general results. In this paper, we chose seven classic algorithms as innovation instances, as shown in table 1.

*Table 1.* A list of algorithm examples and the original article metadata corresponding to each algorithm. Total number of authors is the number of authors extracted from citation of citation after author name disambiguation.

| Algorithm | Original article | Citation (-2018) | Citation of citation (-2018) | Total number of author |
|---|---|---|---|---|
| LSA (Latent Semantic Analysis) | Deerwester, S., Dumais, S. T., Furnas, G. W., Landauer, T. K., & Harshman, R. (1990). Indexing by Latent Semantic Analysis. Journal of the Association for Information Science and Technology, 41(6), 391–407. | 9071 | 214388 | 317442 |
| SVM (Support Vector Machines) | Burges, C. J. C. (1998). A Tutorial on Support Vector Machines for Pattern Recognition. Data Mining and Knowledge Discovery, 2(2), 121–167. | 12846 | 488255 | 492488 |
| NMF (Non-negative Matrix factorization) | Lee, D. D., & Seung, H. S. (1999). Learning the parts of objects by non-negative matrix factorization. Nature, 401(6755), 788–791. | 6763 | 93846 | 192213 |
| PLSA (Probabilistic Latent Semantic Analysis) | Hofmann, T. (1999, August). Probabilistic latent semantic indexing. In Proceedings of the 22nd annual international ACM SIGIR conference on Research and development in information retrieval (pp. 50-57). | 3389 | 73798 | 124314 |
| LDA (Latent Dirichlet Allocation) | Blei, D. M., Ng, A. Y., & Jordan, M. I. (2003). Latent dirichlet allocation. Journal of machine Learning research, 3(Jan), 993-1022. | 17371 | 158638 | 244610 |
| LDA sample | ~10% sample of LDA citation | 1771 | 36787 | 64983 |
| Word2Vec | Mikolov, T., Chen, K., Corrado, G., & Dean, J. (2013). Efficient estimation of word representations in vector space. arXiv preprint arXiv:1301.3781. | 2065 | 35610 | 67037 |
| Glove (Global | Pennington, J., Socher, R., & Manning, C. D. (2014, | 3643 | 14526 | 29751 |

| Vectors for Word Representation) | October). Glove: Global vectors for word representation. In Proceedings of the 2014 conference on empirical methods in natural language processing (EMNLP) (pp. 1532-1543). |

We chose the original literature that proposed these algorithms as the starting point of innovation and their citation as adoption. Since the original literature (Boser et al., 1992) of SVM has been published for a long time, follow-up researchers tend to cite a tutorial paper explaining the usage of SVM. Therefore, we chose the tutorial work (Burges, 1998) for SVM instead.

There are currently three main databases that are viewed as authorities for providing high quality citation data, Web of Science, Scopus and Microsoft Academic Graph (MAG). Here, we use the collection experiment of citation data of LDA to determine which database is more suitable for our analysis. First, we searched citations of the LDA article on July 10, 2020 and the collected citations range from January 1, 2003 to December 31, 2019. As shown in Figure 1, each circle represents a data set, and each circle's size represents the number of citations contained. The numbers on the edges are the duplicated citations between two datasets, which are calculated based on the combination of the title and the author's initial and last name. The results show that MAG (19858) has the largest amount of the citation data, and contains 78% (8531/10938) of Web of Science and 72% (12038/16720) of Scopus data. Due to the different metadata formats, the fusion of the three data sets may cause errors, so we choose MAG as the data source.

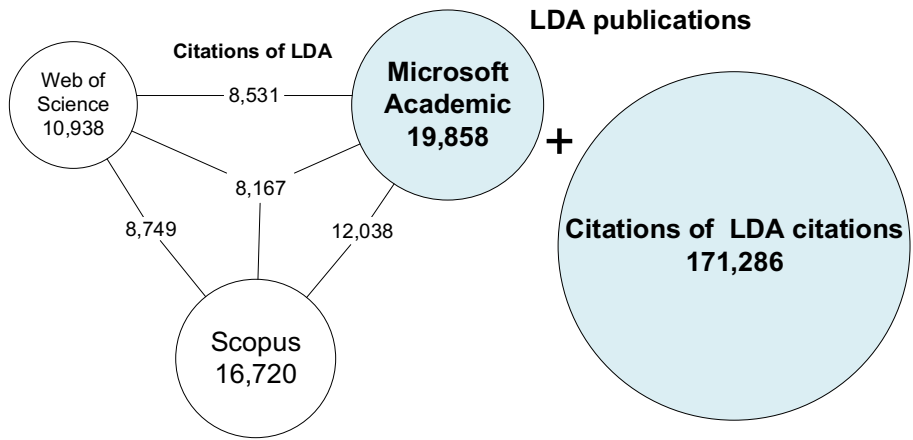

*Figure 1.* The distribution of LDA citations in three databases, the three circles represent the number of citations included in Microsoft Academic Graph (MAG), Web of Science, Scopus, and the edge between them represents the number of duplicated citations.

In order to comprehensively cover the entire process of diffusion, we added the citations of the paper that has cited the algorithms directly to our dataset. we also remove data with missing values (such as title, author, date, etc.). Furthermore, all the publications of the unique authors have been collected from the MAG database. Because there is missing data in the last two years in MAG, we only collected citation data up to December 31, 2018 for each algorithm. At the same time, in order to verify the impact of missing data on the results, we performed a bootstrapping estimation by randomly sampling from the empirical distributions of LDA citations. As such, this random sampling facilitates estimating the statistical significance of the difference between the whole LDA citation and 10% LDA sample.

In the MAG database, each unique author is represented by one or more author IDs. The task of the author disambiguation is to enumerate all duplicate cases for each author ID in a given indexed author ID set. In this paper, we used the system RankMatch proposed in the second track of the 2013 KDD Cup Data Mining Competition, which is dedicated to the author disambiguation task of Microsoft Academic Graph (Liu et al., 2013). After pre-processing, the similarity of paired authors is calculated through a variety of heterogeneous connecting paths such as "author-paper-author" and "author-paper-venue-paper-author", etc. After the author name disambiguation, 100 pairs of merged authors and 100 pairs of authors who were not merged but had the same name were randomly selected and manually judged. The accuracy reached 91% and 94%, respectively.

Defining the author's research domain can help us understand the cross-domain mechanism of the diffusion process. Since most of the authors' affiliations in MAG are missing or without specific departments, it is difficult to determine their research areas. Therefore, we use the fields-of-study (FoS) of all papers published by the authors to determine their major research topic. MAG classifies the research topics of papers into fields-of-study through semantic analysis (Sinha et al., 2015; Effendy & Yap, 2017). The highest level consists of the following 19 fields (in alphabetical order): Art, Biology, Business, Chemistry, Computer Science, Economics, Engineering, Environmental Science, Geography, Geology, History, Materials Science, Mathematics, Medicine, Philosophy, Physics, Political Science, Psychology and Sociology. We collect all papers for each author, count the frequency of each FoS, and select the one with the highest frequency as the author's research domain.

### 3.2 Constructing the diffusion tree

Academic collaboration is one of the most obvious clues and the strongest relationships to knowledge sharing. On the one hand, tracking the diffusion path of an innovation is difficult in reality. According to Rogers (2010), diffusion occurs through a five-step decision-making process: knowledge, persuasion, decision, implementation, and confirmation. There are too many information sources to nail down during the diffusion process and many of them are hard to trace. However, citations document every adoption and provide us the records of the diffusion path through the whole lifetime of an innovation. On the other hand, although an innovation can be communicated in a variety of ways, interpersonal contacts have been found to be critical for building trust and eliminating technological barriers to facilitate the exchange of information about new ideas (Corner & Tran, 2016). Most potential adopters base their judgments of an innovation on information from those who have sound knowledge of it and who can explain its advantages and disadvantages.

Research also shows that scientists are more willing to find, validate, and filter information based on social interactions (Pontis et al., 2017). As early as 1977, Allen (1977) found that engineers and scientists were roughly five times more likely to turn to a person for information than to an impersonal source such as a database or file cabinet. Furthermore, scientists embedded in collaboration networks share ideas, use similar techniques, and otherwise influence each other's work (Owen-Smith, 2001; Yang et al., 2015). As Rogers elaborates, the heart of the diffusion process consists of the modeling and imitation by potential adopters of their network partners who have adopted the innovation previously.

Based on the above analysis, we divide the diffusion channel into two types, broadcasting and virality as shown in Figure 2. The broadcasting diffusion is that an innovation is transmitted directly from the original source to the adopter creating the first layer of the diffusion tree. For example, adopters who cite LDA through the broadcasting channel by reading the original LDA paper or the citations of LDA, listening to lectures, or getting information on the Internet, etc. In this process, LDA is diffused from one to many. While virality is the way in which the adopter learns the innovation from other adopters or collaborators and decides to adopt it. Here, the LDA in the viral diffusion channel is passed from one person to another through scientific collaboration. For example, experienced scientists introduce or teach their collaborators the methods they used before. These collaborators will continue to use the methods later in their own research and pass to others through their further collaborations.

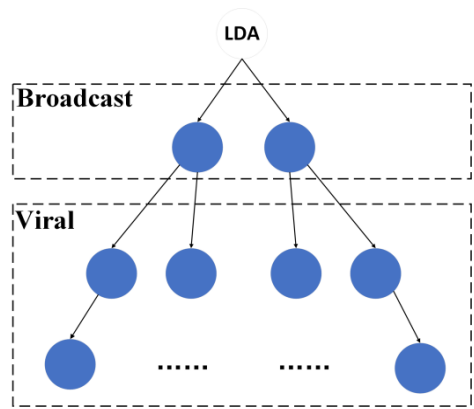

*Figure 2.* A schematic depiction of broadcasting vs. viral diffusion, where nodes represent individual adopters and directed edges indicate who diffused the innovation

to whom.

Therefore, we use the coauthorship as a proxy to the diffusion channel and build a diffusion tree for each algorithm. Below we use LDA as an example to illustrate our process of constructing a diffusion tree:

- First, for each author, we construct a coauthor vector as {A| B, C, D, …} from the whole LDA citation corpus. For each author, we define the adoption time as the publish time of his/her first paper citing LDA or LDA's citations if he/her never cite LDA directly. For example, Time(A) = [20040201], Time(B) = [20101101], Time(C) = [20120405].

- Second, we randomly choose an author B and the coauthorship vector for its neighbors is {A, C, D, ……}. If the adoption time of B is earlier than all of his/her coauthors, then we add a directed link (LDA, B) to the diffusion tree which means B adopt LDA directly. If not, we choose B's coauthor A with the earliest adoption time among all the coauthors and define that LDA is diffused from A to B and we add a directed link to the diffusion tree.

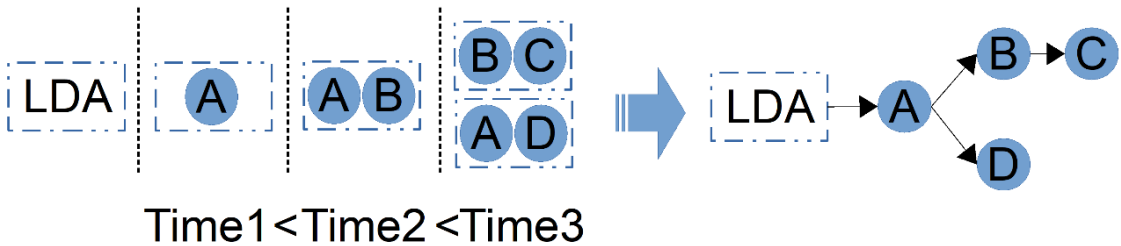

*Figure 3.* An example demonstrating the procedure to construct the diffusion tree. The first dotted square represents the work of LDA and each of the rest dotted squares is one citation of LDA or one citation of the LDA citations. A circle represents an author, and two circles in one square means coauthorship.

### 3.3 Measuring the innovation diffusion

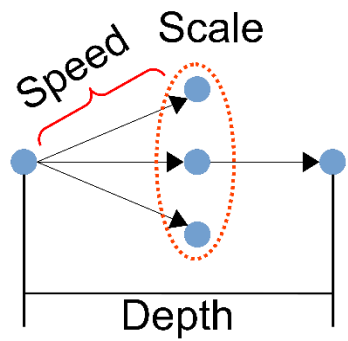

*Figure 4.* Three dimensions of the diffusion tree.

We analyze the functions of diffusion channels in the diffusion tree through three different dimensions, as shown in Figure 4:

- *Scale* is the number of children of one node which can represent its diffusion power. Here, scale measures the number of adopters of each algorithm.

- *Depth* is the number of layers from one node to the leaf of its brunch.

- *Speed* is the interval length of the adoption time between a child node and its parent.

For an adopter x, the interval time $it(x)$ can be calculated by following equations, measured by days using LDA as example:

$$it(x) = \begin{cases} T(x)_{firstLDA} - T(x)_{firstPaper}, & \begin{pmatrix} x \in Broadcasting \\ T(x)_{firstPaper} \geq T(x)_{LDA} \end{pmatrix} \\ T(x)_{firstLDA} - T(x)_{LDA}, & \begin{pmatrix} x \in Broadcasting \\ T(x)_{firstPaper} < T(x)_{LDA} \end{pmatrix} \\ T(x)_{firstLDA} - T(x)_{firstPaper}, & \begin{pmatrix} x \in Virality \\ T(x)_{firstPaper} \geq T(px)_{firstLDA} \end{pmatrix} \\ T(x)_{firstLDA} - T(px)_{firstLDA}, & \begin{pmatrix} x \in Virality \\ T(x)_{firstPaper} < T(px)_{firstLDA} \end{pmatrix} \end{cases}$$

where, $T(x)_{firstPaper}$ is the publication time of $x$'s first paper among all his/her publications. $T(x)_{firstLDA}$ is the publication time of $x$'s first paper citing LDA or LDA's citations if x never cite LDA directly. $T(x)_{LDA}$ is the publication time of LDA or the one of LDA's citations that $x$ cited earliest if he/she never cite LDA directly.

For broadcasting members, if they start their research career after the publication of LDA or the LDA citation they cited earliest, the interval time $it(x)$ is the time between $T(x)_{firstLDA}$ and $T(x)_{firstPaper}$. Else if they start their research career earlier, the interval time $it(x)$ is the time between $T(x)_{firstLDA}$ and $T(x)_{LDA}$.

For virality members, if they start their research career after the adoption time of their parents, the interval time $it(x)$ is the time between $T(x)_{firstLDA}$ and $T(x)_{firstPaper}$. Else if they start their research career earlier than the adoption time of their parents, the interval time $it(x)$ is the time between $T(x)_{firstLDA}$ and $T(px)_{firstLDA}$.

The shorter the interval time, the faster the diffusion speed. Therefore, we define diffusion speed $DS(x)$ as follows:

$$DS(x) = \frac{1}{it(x)/365 + 1} \quad (0 < DS(x) \leq 1)$$

## 4. Results

## 4.1 Diffusion scale

To quantify the diffusion power of each channel, we need to construct the spreading trajectory for each innovation. Based on the method of tracking innovation diffusion as defined above, we constructed the diffusion tree for each algorithm. Here, we only visualize the diffusion tree constructed with LDA citations. As shown in Figure 5, we use Gephi (Bastian et al., 2009) to construct the diffusion tree and apply ForceAtlas2 layout algorithm (Jacomy et al., 2014, p. 2) to map the structure. It presents a top view of a canopy, where the root node LDA located in the center, and the other adopters grow outwardly layer by layer.

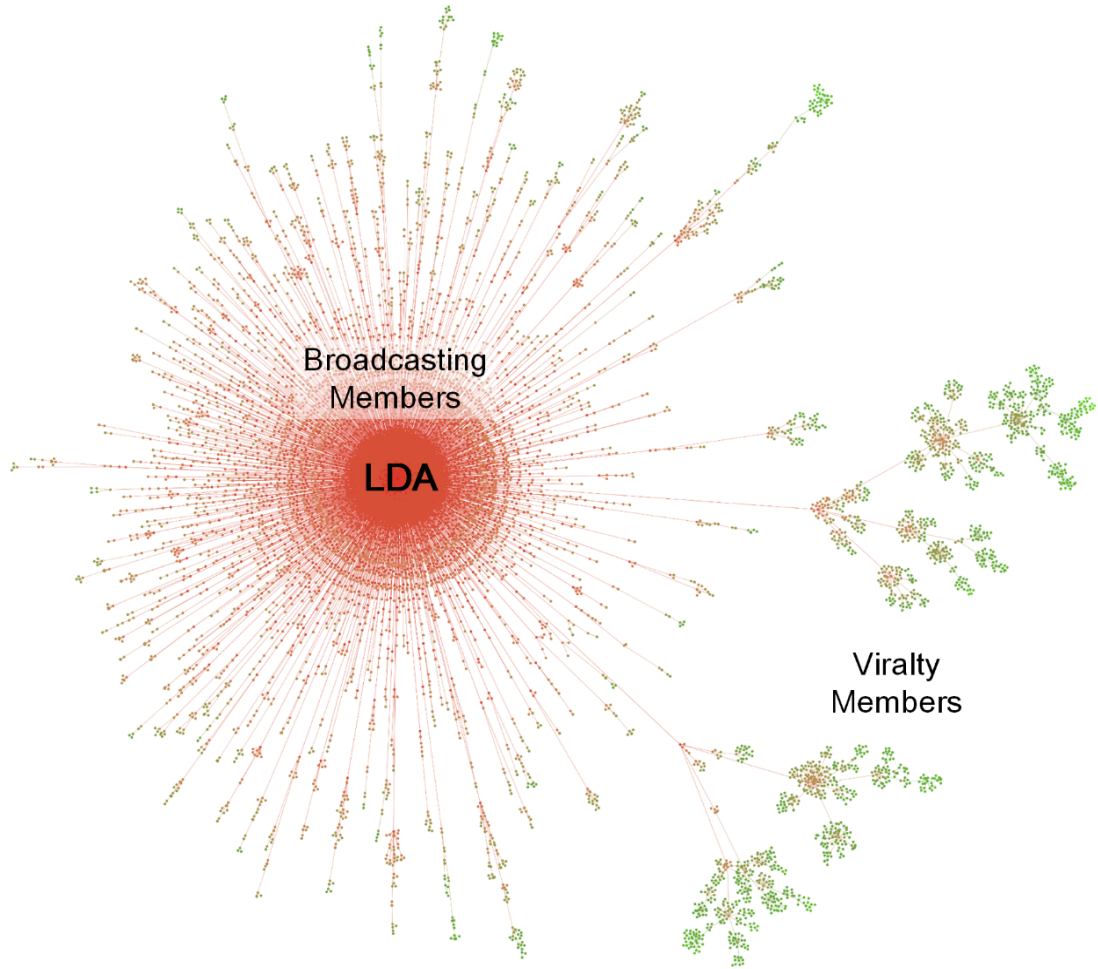

*Figure 5.* The diffusion tree of LDA (only with LDA citations). The nodes in the center surrounding LDA (root node) are the broadcasting members, which are all directly connected to the tree root and build up the first layer. The vitality members constitute the branches and leaves of the tree and extend outward. The nodes gradually change from red to green from the inside to the outside.

Here, we refer to the nodes directly connected to the LDA as broadcasting members, and the other ones as virality members. As can be seen from Figure 5, LDA is densely surrounded by a large inner ring grouped by broadcasting members and the virality members are constantly stretching outwards to form a lot of branches.

Over the past 18 years, the citation of each algorithm has experienced rapid growth. Figure 6 shows the annual number of new citations in the diffusion process of every algorithm. There are three type of diffusion patterns. LSA, SVM, NMF, and PLSA in the elderly group have gradually declined after long-term growth. The growth rate of the middle-aged algorithm LDA has slowed down and gradually reached its peak. Due to the development of natural language processing and deep learning technology, new language models based on word embedding (Pouriyeh et al., 2018) or hierarchical network (Hwang & Sung, 2017) are proposed recently to replace LDA for topic modeling. The youth Word2Vec and Glove are in an explosion stage. It shows that in recent years, the take-off stage of the innovation diffusion has become shorter and faster, which corresponds to the fading speed of traditional algorithms.

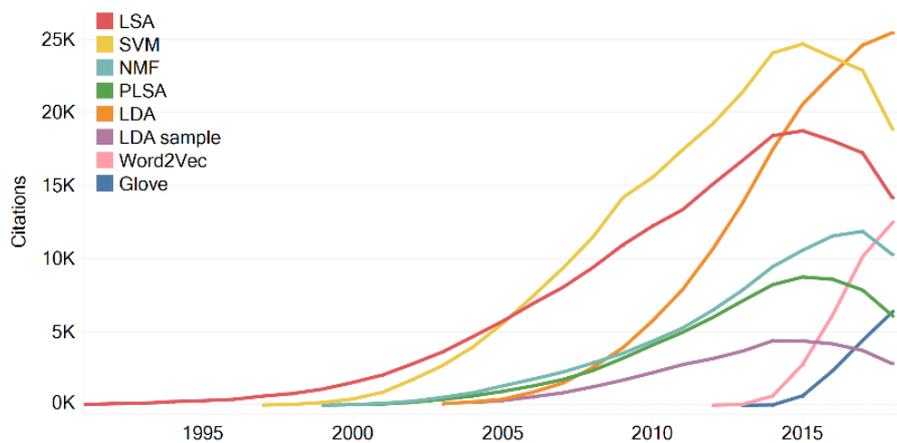

*Figure 6.* The yearly distribution of new citations in the diffusion process of each algorithm.

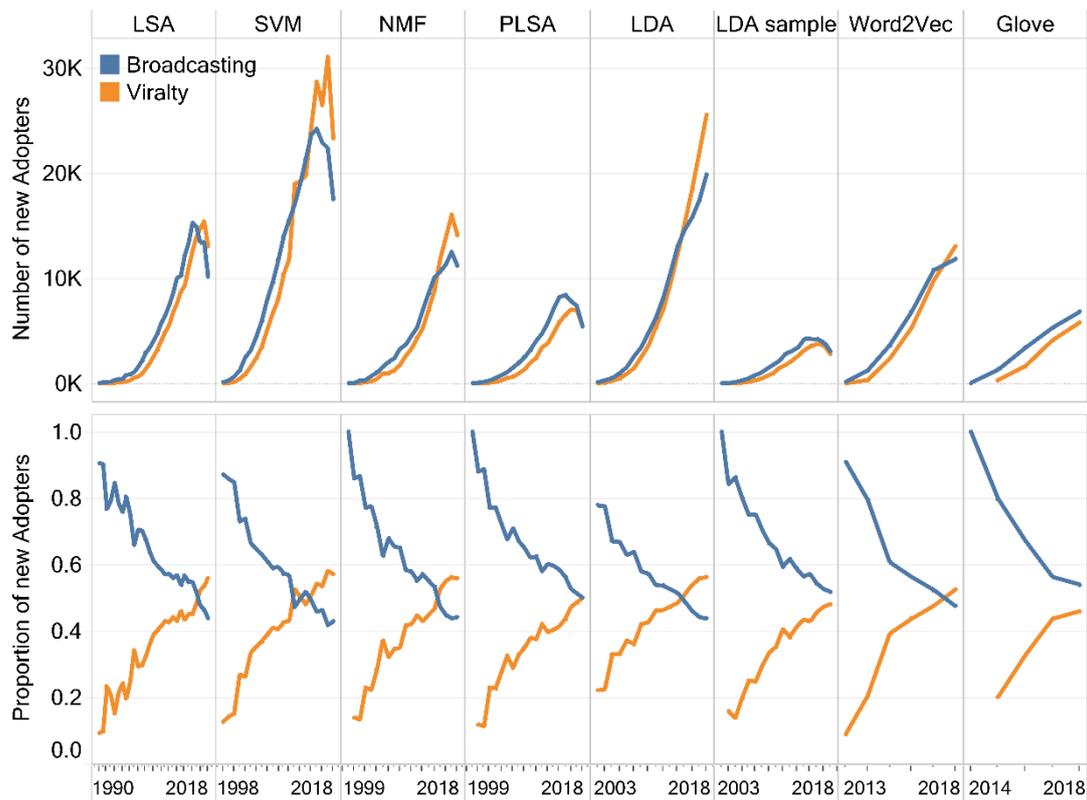

*Figure 7.* The yearly distribution of new broadcasting and virality members and proportion distribution of different types of adopters.

Both virality and broadcasting play a highly important role and increased simultaneously in the diffusion process. Broadcasting is more powerful in the initial stage, as shown in Figure 7, but may be gradually caught up or even surpassed by virality in the later period. When a study is completed, a paper will be published to report the innovation. At this moment, downloading and reading the article is the most representative way of broadcasting. As the publication gets older, due to the fast growth of published articles and diverse ranking mechanisms of search engines, the probability of encountering the original paper in the database or on the Web is reduced.

The diffusion process of the algorithms all conform to the S curve of the technology adoption life cycle. Most of the innovations have gradually become the state-of-the-art algorithms, for example, LDA is sometimes even regarded as synonymous with topic modeling due to its excellent scalability and easily interpretable results (C. Wang & Blei, 2011). Furthermore, the technical threshold for adopting computer algorithms is constantly dropping and interdisciplinary teamwork is increasing. With the development of open source computer communities such as

GitHub, popular algorithms can basically find support tools that can be adopted quickly. Some may be the original code written by the authors, and some may be developed specifically for scholars from other fields (Cohen Priva & Austerweil, 2015).

### 4.2  Diffusion depth

The diffusion depth mainly explains the role of viral diffusion. In this section, we first analyze the depth and structural virality of the diffusion tree and then explore how cross-domain diffusion related to the diffusion depth.

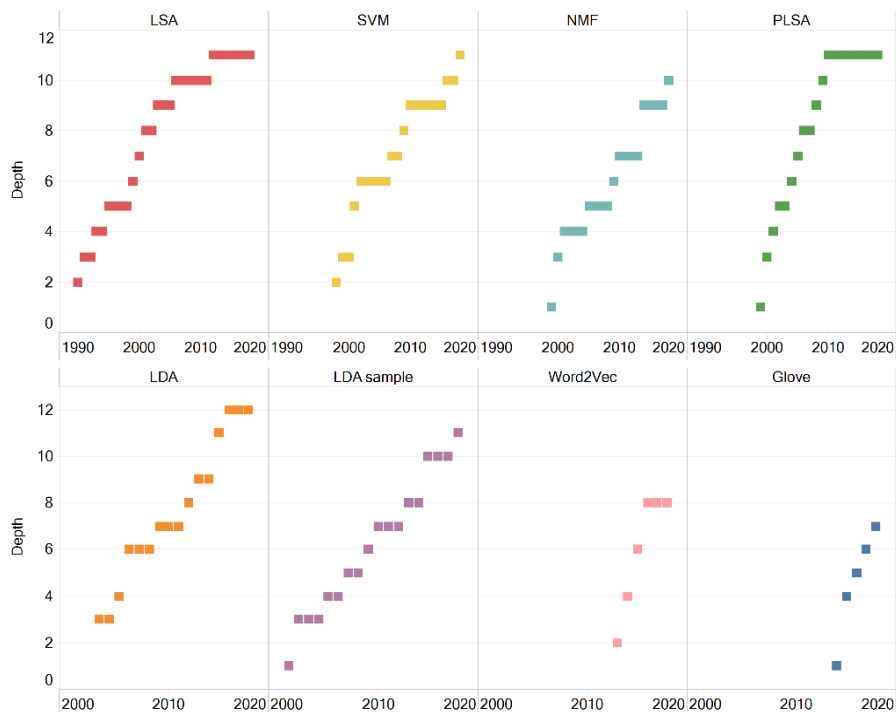

*Figure 8.* The depth of the diffusion trees.

Figure 8 shows the diffusion depth of the tree. We can see that the depth of most trees is 10 or 11, LDA reaches 12 layers at most, Word2Vec and Glove are relatively young, so there are only 7 or 8 layers. The trees will get new growth space when leaf nodes continue to transfer the innovations to others and become branch nodes. However, in recent years, the trees are no longer growing deeper. This phenomenon may be driven by multiple reasons. First, an innovation can gradually lose its heat over time, and the number of adopters will gradually decrease. Second, scholars working in one particular field is limited, so the number of adopters will not grow indefinitely. Third, new and better innovations will replace the current innovation. When a certain depth is reached, scholars in the field would have already known the innovation and no longer need to learn from others.

In addition, the academic community is relatively small. An early study found that the average distance of the scientific co-authorship network has been stabilized to about 6 for years (Elmacioglu & Lee, 2005). Since we are using the algorithms' citations and the citations of the paper that has cited the algorithm directly, the longest path in the tree graph can be regarded as twice that of a six-degree separation(Newman, 2001).

Next, we define two indicators, activation rate and growth rate, to measure the depth growth pattern of the diffusion tree. Here, active members refer to the members who can continue to pass the innovation to others after adoption, and the activation rate is the proportion of active members in the total adopters of the #n layer. The growth rate indicates that the number of adopters in #n layer divided by the number of active members in the previous(#n-1) layer.

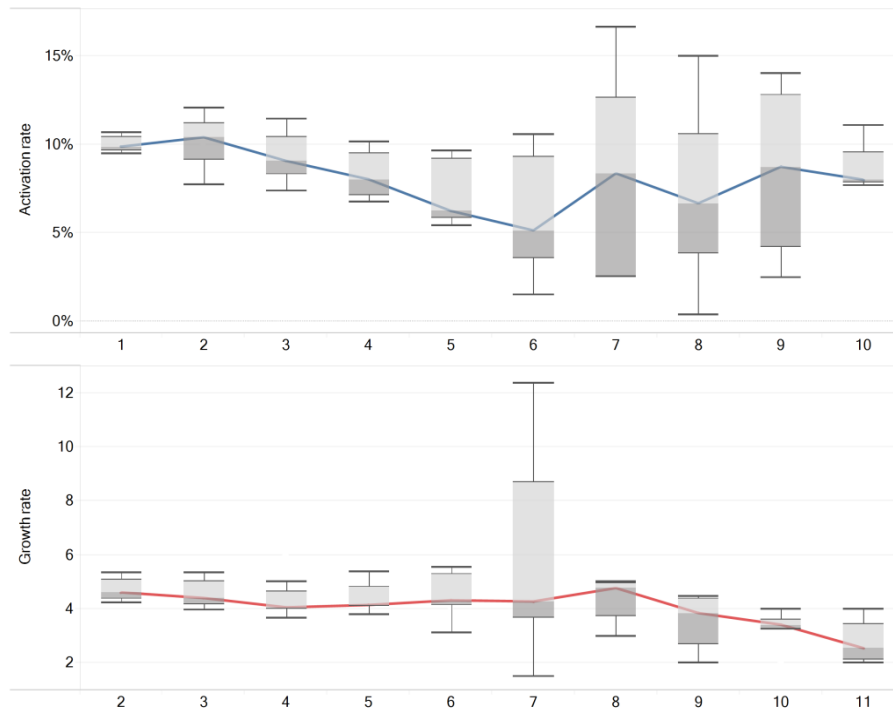

*Figure 9.* Activation rate and growth rate in each layer. The X axis indicates the number of layers. We calculate the activation rate and growth rate of each algorithm (including LDA sample) in each layer, and then use box plots to show their distribution. The lines in the figure connect the medians of each layer.

As the layers increases, the number of adopters is decreasing. The activation rate of each layer starts from 10% and gradually decreased. Except for the last three layers (the number of members is too small to be considered), the growth rate for each layer remains approximately 5. In other words, for every layer in the diffusion tree, less than 10% of nodes will continue to grow and get 5 times more nodes for the next layer.

To gain a further understanding of the meaning of diffusion depth, we analyze the research domain distribution of adopters at different diffusion layers (see Figure 10).

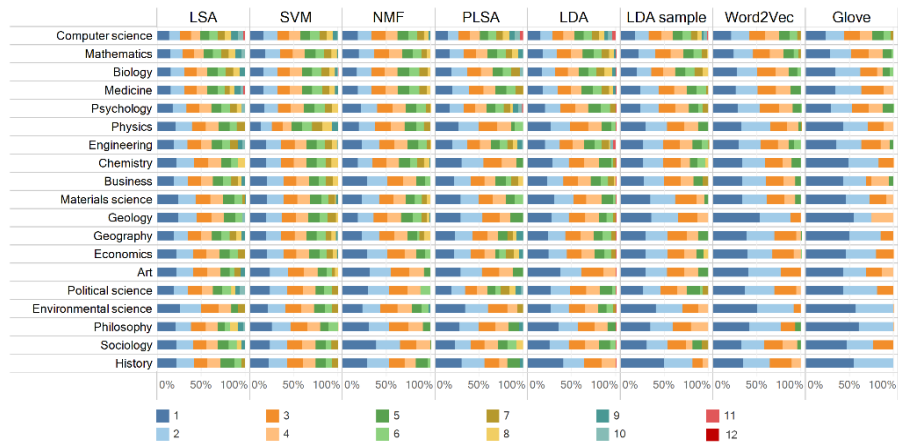

*Figure 10.* Proportion distribution of scientists from different domains at different diffusion layers (Notes: Since the number of scholars in computer science differs greatly from those in other domains, we take a log on the number of adopters). 1-12 represents different layers and is distinguished by different colors.

The more relevant the domain is associated with the innovation, the more likely it is to spread deeply, and vice versa, the more distant domains are spread shallowly. Our algorithm cases are all proposed in computer science. Dias et al. (2018) studied the similarity between scientific disciplines based on expert classification, citation and language use. From the results of dissimilarity and clusters of disciplines, they found that computer science is closely related to mathematics and engineering, and far from art and philosophy. The distance between the various disciplines in Figure 10 is similar to the results of their study. As can be seen from Figure 10, only the diffusion of LDA in computer science and engineering have penetrated the 11th layer, while most of the others stay between the first and fifth layers. History and art, whose subjects are distant from the computer science, only reach layer seven at most. From this point, we can see that cross-domain diffusion often occurs in the broadcasting channel (the first layer), and the initial stage of the viral diffusion (i.e., the first four layers).

Correspondingly, researchers in other disciplines may read the original article or the citations of the algorithm, and computer scientists may collaborate with researchers from other disciplines and pass different algorithms to them. For example, in a recent study (Yoshida et al., 2018) about the evolution of whole-rock composition during metamorphism, LDA was introduced into geology and used to find endmembers based on the frequencies of elements that make up the rock of interest. Most of the authors, Yoshida, Kuwatani, Hirajima and Iwamori, are scientists from geology or earth science, while one of them, Shotaro Akaho, is from computer science. Also, the growth of innovation requires soil that matches itself, which is happening in the field of computer science, then can be carried further by other scholars through layer-by-layer collaboration. Computer scientists are continuously upgrading LDA and proposing other modified models, such as Labeled LDA (Ramage et al., 2009) and TM-LDA (Wang, Agichtein, & Benzi, 2012, p.123). Some of these new models have also been cited thousands of times and diffused to other domains.

### 4.3 Diffusion speed

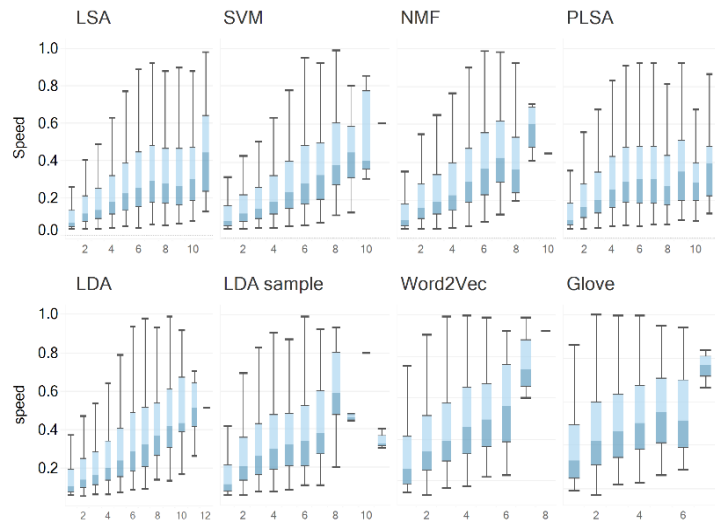

*Figure 11.* The diffusion speed in different layers. The first layer represents the diffusion speed of broadcasting members, and the other layers represent virality members.

For the diffusion speed at the individual level, the virality is significantly faster than the broadcasting, as shown in Figure 11. With the increase of the number of layers, the diffusion speed is generally getting faster. On the one hand, virality is achieved through collaborating with experienced scientists who have already adopted the innovation. Working with experienced adaptors can get to know the innovation faster than learning alone, especially for scientists from a distant field. Meanwhile, when a potential adopter observes innovation benefits, expected risk decreases and the likelihood of adoption increases (Wejnert, 2002). That means when a viral member learned the innovation via his/her collaborators who have already tried the innovation, the probability for a viral member to adopt the innovation is much higher. On the other hand, combined with the results in the previous section, the deep diffusion mainly occurs in computer science and its similar domains. Therefore, the speed of diffusion will be accelerated if both the distributors and receivers are from related domains and understand/ absorb each other's idea quickly.

Since virality plays an important role in the diffusion process, will it affect the future behavior of the adopter? For different types of adopters, we calculate the proportion of one author's articles citing the algorithm to one's total publications. This can help us answer the question which types of authors are more willing to adopt the innovation repeatedly.

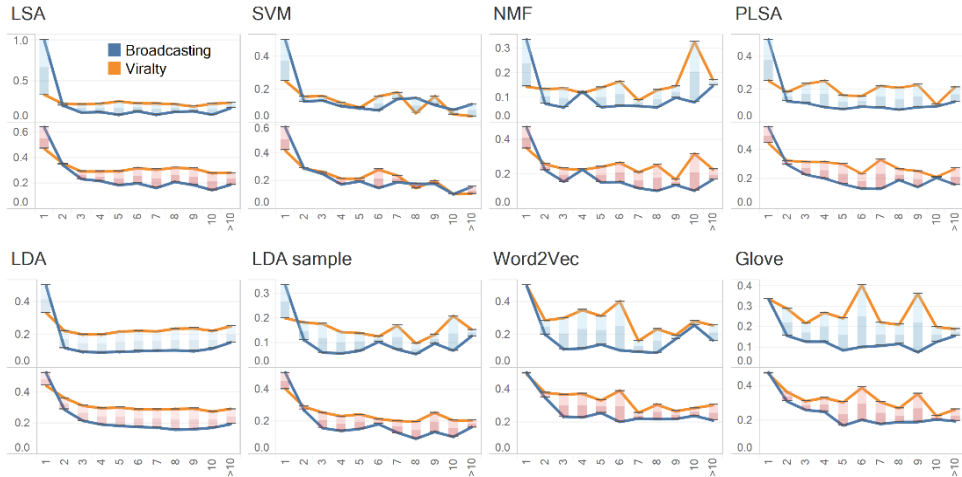

*Figure 12.* The proportion of one's algorithm citations in his/her total number of publications (>n) for different types of adopters (%). The X-axis represents the total number of papers published by a scholar citing the corresponding algorithm, the upper half of the Y-axis (blue) represents the median line of the proportion of algorithm papers, and the lower half (red) is the average line of the proportions. Box diagrams are used to illustrate the differences between the two groups.

Compared with broadcasting members, virality members tend to cite the innovation more times and this acquired innovation will become an important part of their research practice. For authors who cited the corresponding algorithm twice or more, virality members are more willing to recite it than broadcasting members. This difference can be interpreted by the difference of viral and broadcasting members. Rogers (2010) classified adopters into five different segments as innovators, early adopters, early majority, late majority, and laggards. Early adopters like to try new innovations, take risks, but late majority and laggards like to get innovation tested by others through interpersonal communication. This keeps them stay longer on a certain innovation.

## 5. Conclusion

This article proposes a method of building an innovation diffusion tree based on scientific collaboration and explore broadcasting channel and virality channel of innovation diffusion from three aspects: scale, depth and speed.

Results show that: (1) Broadcasting and virality have the same power but play different roles in promoting the diffusion of innovation. The growth of new adopters in two different channels is basically the same indicating the same power in promoting the diffusion scale. Broadcasting plays a major role in the early stage, while the influence of virality is increases gradually over time.

(2) Virality is the driving force of increasing layers of diffusion, which means that researchers continue to adopt innovation through cooperation, and their identities can transform from recipient to promoter. We find that 10% of the researchers at the first layer will continue to disseminate innovation to others and the ratio will gradually decrease as the diffusion goes deeper. Meanwhile, each active node in each layer will pass the innovation on to about 5 people. This simple incremental relationship captures the growth characteristics of the diffusion tree well, which indicating the diffusion process follows a random but repeatable process.

(3) Virality is faster for the first adoption of adopters. Rogers (2010) indicated that while mass communication is an effective communicative mode, interpersonal communication is a direct, double-sided, selective, and reciprocal mode. In terms of the growth rate of new adopters, broadcasting is more powerful in promoting the diffusion than virality. While focusing on the first adoption, the speed between adopters in virality is significantly faster than that in broadcasting, and researchers from virality channels tend to adopt the innovation multiple times.

We took a random sample from the largest LDA data and analyzed it simultaneously with other algorithm data. The results showed that missing data did not affect the consistency and continuity of the findings. This research also faces some limitations. The data of the diffusion tree comes from MAG which does not cover all scientific publications and our innovation cases are all computer algorithms, so the generalization of the conclusions needs further verification. In addition, there are other informal and formal channels of innovation diffusion in the scientific community, most of which cannot be recorded in citing and coauthoring behaviors. Interpreting coauthorship as the virality channel can be inadequate, as scholars can still communicate with each other without co-authoring a paper together.

In the future, we will continue to explore innovation adoption through the diffusion process of scientific innovation. Using content-based citation analysis, we can restore the comprehensive diffusion process and identify the function of the innovation for the adopters. Furthermore, an important question waiting for an answer is how an innovation diffuses in different fields and breaks the potential barriers between various domains.

## ACKNOWLEDGEMENTS

This work is supported by National Social Science Fund of China (No. 18CTQ027). The authors are grateful to all the anonymous reviewers for their precious comments and suggestions.